\documentclass{l4dc2027}

\usepackage{graphicx}
\usepackage{caption}

\usepackage{tikz}
\usepackage{amsmath,amssymb}
\usetikzlibrary{shapes,arrows.meta,positioning}

\title[Online Neural Network Training via Closed-Loop Model-Free Tracking Control]{Online Neural Network Training via Closed-Loop Model-Free Tracking Control: An Empirical Study}
\usepackage{times}

\author{%
 \Name{Loïc MICHEL} \Email{loic.michel@ec-nantes.fr}\\
 \addr Nantes Université, École Centrale Nantes, CNRS, LS2N, UMR 6004, 44000, Nantes, France.
 \AND
 \Name{Jean-Pierre BARBOT} \Email{barbot@ensea.fr}\\
 \addr Nantes Université, École Centrale Nantes, CNRS, LS2N, UMR 6004, 44000, Nantes, France. \\
 \addr Quartz Laboratory EA 7393, ENSEA, 95014 Cergy-Pontoise Cedex, France.
}

\begin{document}

\maketitle

\begin{abstract}%
Training artificial neural networks predominantly relies on offline, gradient-based backpropagation algorithms. In contrast, biological circuits process and retain information through decentralized, continuous-time  adaptation. In this work, we explore an online, gradient-free learning methodology by formulating the tuning of synaptic weights as a closed-loop trajectory tracking problem driven by model-free control principles. Rather than performing global optimization across static parameters, decentralized controllers adjust synaptic interactions in real-time using solely local input-output tracking errors. We present an exploratory benchmark neural architecture, whose purpose is to assess the dynamic resilience of the proposed approach, particularly in tracking target outputs and adapting to abrupt non-stationarities and structural perturbations (such as topological changes or dynamic dropout) without requiring offline retraining. This empirical preliminary study provides a proof-of-concept for control-theoretic learning paradigms and outlines the foundation for scalable, dynamically evolving connectome frameworks.
\end{abstract}


\section{Introduction}

Training deep neural networks in machine learning and data science relies predominantly on backpropagation and gradient-based optimization algorithms.  By viewing a neural network as an interconnected dynamical system whose synaptic connections evolve in real-time, control-systems approaches offer promising alternatives for online learning. 
\noindent
Perspectives in structural plasticity have highlighted the necessity of developmental architectures that autonomously evolve their topologies \cite{Kowaliw2014Growing}. More recently, framing deep residual architectures through nonlinear control theory has formally established their universal approximation capabilities \cite{Tabuada}. 

Both control theory and machine learning aim at modeling, extracting knowledge, and performing decision-making in complex environments. In parallel, closed-loop control principles have been increasingly recognized as essential mechanisms governing both biological neural circuits and neural interfaces. 
Furthermore, state-space representations have become a cornerstone in neural engineering to dissociate intrinsic recurrent dynamics from exogenous task inputs \cite{Shanechi}, while closed-loop adaptive decoders have been shown to actively guide functional plasticity under unmodeled recording drift \cite{ORSBORN}. Biologically plausible alternatives have actively sought to bypass global backpropagation. Notably, \cite{lillicrap} demonstrated that random synaptic feedback weights suffice to transmit error signals and support effective learning, and the Direct Feedback Alignment strategies \cite{Nokland2016DirectFA} have demonstrated that error signals can be projected directly to individual hidden layers.

From an architectural perspective, continuous-time formulations such as Neural Ordinary Differential Equations \cite{chen2018neural} and Liquid Time-Constant networks \cite{Hasani2021LTC} have highlighted the benefits of endowing hidden states and synaptic conductances with continuous temporal dynamics. While these architectures are usually optimized through adjoint-state or gradient-based methods across temporal horizons, exploring alternatives like closed-loop learning-based strategies remains a fertile and complementary direction. 

Recently, a growing body of literature has established bridges between control theory and machine learning, highlighting structural equivalences between adaptive control and optimization methods. Combining these domains opens rich perspectives, see e.g. the survey \cite{Cerf}: adaptive control strategies especially can be coupled with machine learning architectures to enhance convergence under time-varying parameters, unmodeled dynamics, and non-stationary input distributions \cite{Gaudio}, and deep-learning can be cast as continuous-time optimal control problems governed by Pontryagin's maximum principle \cite{E2019MeanField}, bridging deep architectures and variational trajectory optimization. Furthermore, foundational connections between self-tuning regulators and reinforcement learning algorithms have formalized the convergence of model-free methods toward optimal control in unknown systems \cite{Matni}. 

In parallel, network control theory has been successfully deployed to study biological connectomes from a dynamical perspective, as demonstrated by \cite{Morrison1, Morrison2} in the nonlinear control of interconnected neural circuits. Concurrently, nonlinear contraction theory has emerged as a fundamental tool to guarantee exponential stability and convergence in interconnected recurrent networks under structural parameter shifts \cite{Davydov2022Contractivity}. In \cite{Parkes2024NCT}, linear state-space models and controllability Gramians characterize the energy required for brain-state transitions. 

This interaction between feedback adaptation and structural resilience provides strong motivation for exploring decentralized, gradient-free control laws directly within artificial neural architectures. Building upon these insights, our work explores an intrinsic perspective: endowing synaptic connections with their own internal control dynamics to achieve decentralized and  real-time learning. This work formulates the supervised training of an artificial neural network directly as a closed-loop trajectory tracking problem. Rather than relying on gradient backpropagation, we employ a {\it model-free and derivative-free control} strategy to dynamically adapt the synaptic weights in real-time. We conceptualize the network as a dynamical system where the connections themselves possess intrinsic state dynamics, enabling  structural plasticity. Through numerical simulations, we
show how it could be possible that closed-loop controllers allow neurons to track target outputs, adapt online to abrupt structural changes (such as  connection dropouts), and exhibit memory retention without requiring offline retraining.

The proposed approach leverages the model-free control framework, initially introduced in the seminal work of \cite{Fliess2013}, which permits the control of unknown nonlinear systems using only input-output  data. We extend this principle using the "Para-model" control formulation proposed in \cite{michel2018para} by the first author\footnote{See e.g. recent applications in HIV epidemiology \cite{michelHIV}, 
CFD-based Navier-Stokes \cite{ECC_ISIS} and wind turbines control \cite{michelOpenFAST}.}. By eliminating the need for numerical differentiation and explicit model identification, this controller naturally aligns with the universal structure proposed by \cite{Khalil} and endows growing, self-organized neural topologies with promising closed-loop learning capabilities.

The remainder of this paper is organized as follows. Section 2 reviews the  formulation of the "Para-model" control law. Section 3 illustrates the distributed tracking capability on a simple neural network benchmark. Section 4 details the  \texttt{NetPMC} Framework as a dynamic growing connectome. Finally, Section 5 concludes the paper and discusses future directions.

\section{Principle of the model-free and derivative-free control strategy}

Consider a synaptic connection between two nodes modeled as a single-input single-output (SISO) continuous-time dynamical system:
\begin{equation}\label{eq:gen_sys}
\left\{ \begin{array}{l}
\dot{\boldsymbol{x}}(t) = f(\boldsymbol{x}(t), u(t)), \\
y(t) = g(\boldsymbol{x}(t)),
\end{array} \right.
\end{equation}
where $f$ governs the internal synaptic state dynamics with state vector $\boldsymbol{x} \in \mathbb{R}^n$, $u \in \mathbb{R}$ is the local control input, and $y \in \mathbb{R}$ is the observed node output.

To achieve "robust" set-point regulation without identifying the internal vector field $f$, we discretize time with a constant sampling period $h > 0$, letting $k \in \mathbb{N}$ denote the discrete iteration index. The discrete-time "Para-model" control law $\mathcal{C}_{\pi}$ is formulated as:
\begin{equation}
\label{eq:control}
u_k = \Psi_k \sum_{j = 1}^k K_i \, (y^\ast_j - y_j) \, h,
\end{equation}
where $y^\ast_k$ is the reference target, $y_k$ is the controlled output, $K_i > 0$ is an integral gain, and $\Psi_k$ is an online adaptive modulation sequence updating the effective integral computation:
\begin{equation}
\label{eq:ts}
\Psi_k = \Psi_{k-1} + K_p \left( K_\alpha e^{-K_\beta k} - y_k \right).
\end{equation}
The parameter set of $\mathcal{C}_\pi$ consists of positive scalar gains $\{K_p, K_i, K_\alpha, K_\beta\}$. 

A distinctive feature of the proposed controller~\eqref{eq:control} is its pure integral structure. By dispensing with numerical differentiation of noisy tracking signals, the controller filters high-frequency measurement fluctuations, remaining sensitive primarily to the mean tracking error. In~\eqref{eq:ts}, the  recurrence on $\Psi_{k}$ endows the loop with an adaptive memory, while the decaying exponential term $K_\alpha e^{-K_\beta k}$ provides transient excitation before fading asymptotically, mirroring the properties of robust universal integral regulators for relative degree-one dynamics \cite{Khalil}. The overall closed-loop architecture is schematically illustrated in Fig.~\ref{fig:CSM_gen}.

\begin{figure}[t]
\begin{center}
\begin{tikzpicture}[
    auto,
    node distance=2.2cm,
    >=Latex,
    line width=1.3pt,
    draw=blue!65!black,
    block/.style={
        rectangle, 
        draw=blue!65!black, 
        minimum height=1.2cm, 
        minimum width=1.6cm, 
        align=center
    },
    sum/.style={
        circle, 
        draw=blue!65!black, 
        minimum size=1.3cm
    }
]
    \node[sum] (sum) {};
    \node[block, right=1.6cm of sum] (controller) {$\mathcal{C}_\pi(y, y^*)$};
    \node[block, right=2.2cm of controller, draw=black] (system) {$\dot{\mathbf{x}} = f(\mathbf{x}, u)$};

    \node at (sum.west) [shift={(0.22, 0.15)}] {$+$};
    \node at (sum.south) [shift={(0.15, 0.22)}] {$-$};

    \draw[->] ([xshift=-1.6cm]sum.west) -- node[pos=0.1, below=2pt] {$y^*$} coordinate[pos=0.4] (pt_ystar) (sum.west);

    \draw[->] (sum.east) -- node[above] {$\varepsilon$} (controller.west);

    \draw[->] (controller.east) -- node[above] {$u$} (system.west);

    \draw[->] (system.east) -- ++(0.6,0) 
        node[pos=0.4, above] {$y$}
        |- ([yshift=-1.1cm]controller.south) 
        coordinate[pos=0.5] (pt_feedback)
        -| (sum.south);

    \draw[dashed, ->] (pt_ystar) |- ([yshift=1.1cm]controller.north) -- (controller.north);
    \draw[dashed, ->] (pt_feedback -| controller.south) -- (controller.south);

\end{tikzpicture}
\end{center}
\caption{Proposed decentralized "Para-model" control architecture.}
\label{fig:CSM_gen}
\end{figure}

\subsection{Illustrative Example: Decentralized Inversion of Coupled Linear Systems}

To provide intuition on how decentralized model-free controllers coordinate across coupled interconnections without global parameter estimation, consider, for instance, the classical linear system to solve:
\begin{equation}
\label{eq:sys_linear}
\begin{pmatrix}
3 & 0.5 & 8 \\
4 & 7 & 4.5\\
1 & 9 & 3
\end{pmatrix}    
\begin{pmatrix}
x_1 \\
x_2 \\
x_3
\end{pmatrix} =
\begin{pmatrix}
7.95 \\
6.30 \\
3.80
\end{pmatrix},
\end{equation}
whose exact algebraic solution is denoted by $\boldsymbol{x}^* = (x_1^*, x_2^*, x_3^*)^T$. 

We reframe this basic algebraic inversion problem as a dynamic output tracking task for the multi-channel mapping $\boldsymbol{x} \mapsto \boldsymbol{y} = \boldsymbol{A} \boldsymbol{x}$. The objective is to steer the coupled outputs $\boldsymbol{y}(k)$ toward the target vector $\boldsymbol{y}^* = \boldsymbol{b}$, such that $\|\boldsymbol{y}(k) - \boldsymbol{b}\| \to 0$ as $k \to \infty$. 
\noindent
Rather than inverting the matrix $\boldsymbol{A}$  or performing e.g. gradient descent, each decision variable $x_j$ ($j = 1, 2, 3$) is driven by an autonomous, decoupled controller $\mathcal{C}_{\pi j}$:
\begin{equation}
\label{eq:matrix}
x_j(k) = \mathcal{C}_{\pi j}\big(y_j(k), \, b_j\big),
\end{equation}
tuned via its individual parameter set $\{K_{pj}, K_{ij}, K_{\alpha j}, K_{\beta j}\}$. In this formulation, each controller $\mathcal{C}_{\pi j}$ treats the cross-coupling terms $\sum_{l \neq j} A_{jl} x_l$ as unmodeled disturbances and compensates for them exclusively through local output feedback. 

\begin{figure}[t]
  \begin{center}
    {\includegraphics[width=12.5cm]{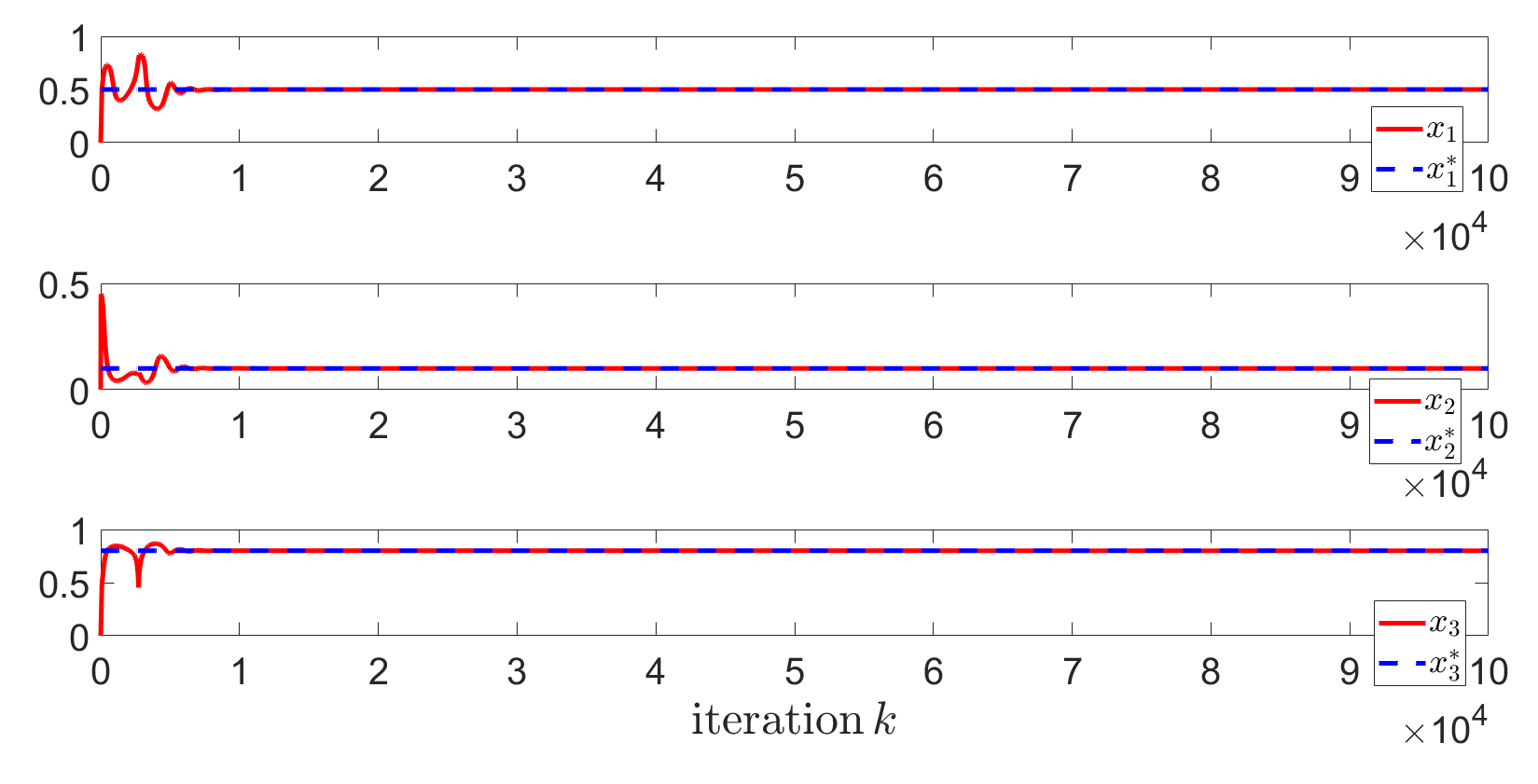}}
    \caption{Evolution of the controlled vector $\boldsymbol{x}$ versus the iterations.}
    \label{fig:fig_result_matrix}
  \end{center}
\end{figure}

As depicted in Fig.~\ref{fig:fig_result_matrix}, the decentralized control signals $\boldsymbol{x}(k)$ converge rapidly to the exact solution $\boldsymbol{x}^*$, illustrating the capability of purely local model-free tracking laws to coordinate and absorb strong algebraic couplings.

\section{Benchmark Connectome Architecture and Online Adaptation}

To assess the feasibility of decentralized model-free tracking within an interconnected neural circuit, we construct an exploratory three-neuron benchmark architecture. The code reproducing the 3-node benchmark simulations and the decentralized $Ax=b$ inversion is made openly available\footnote{
\url{https://github.com/LoicMichelControl/NN\_modelFreeControl}}. Rather than relying on offline gradient backpropagation, the network is formulated as a closed-loop dynamical system where synaptic pathways possess their own internal first-order dynamics and local adaptation controllers.

\subsection{Connectome Dynamics and Broadcast Error Tracking Formulation}

The architecture comprises two sensory inputs ($I_1, I_2$), two hidden neurons ($N_1, N_2$), and one output neuron ($N_3$). Figure \ref{fig:3neuron_arch} depicts the scheme of the 3-neuron network. At discrete step $k$, the neural activation state $\mathbf{z}(k) = [z_1(k), z_2(k), z_3(k)]^T \in \mathbb{R}^3$ evolves according to:
\begin{equation}
\label{eq:node_dynamics}
\begin{aligned}
z_1(k) &= \tanh\big(W_1(k) I_1(k) + W_2(k) I_2(k) + W_7(k) z_1(k-1)\big), \\
z_2(k) &= \tanh\big(W_3(k) I_1(k) + W_4(k) I_2(k) \big), \\
z_3(k) &= \tanh\big(W_5(k) z_1(k) + W_6(k) z_2(k) \big),
\end{aligned}
\end{equation}
where $W_m(k)$ ($m \in \{1,\dots,7\}$) denote the time-varying synaptic weights. For notational compactness, let $\theta_m(k) = \{W_1,\dots,W_7\}$ ($m \in \{1,\dots,7\}$) index the complete set of trainable parameters.

\begin{figure}[t!]
\centering
\begin{tikzpicture}[
    >=Latex,
    line width=1.1pt,
    input/.style={circle, draw=black, fill=gray!15, minimum size=0.8cm, font=\small},
    neuron/.style={circle, draw=blue!65!black, fill=blue!8, minimum size=0.95cm, font=\small},
    weight/.style={font=\footnotesize, inner sep=1.5pt, fill=white}
]

    \node[input] (I1) at (0, 1.3) {$I_1$};
    \node[input] (I2) at (0, -1.3) {$I_2$};

    \node[neuron] (N1) at (3.2, 1.3) {$N_1$};
    \node[neuron] (N2) at (3.2, -1.3) {$N_2$};

    \node[neuron] (N3) at (6.4, 0) {$N_3$};

    \draw[->] (I1) -- node[weight, above] {$W_1$} (N1);
    \draw[->] (I1) -- node[weight, pos=0.28, above right] {$W_3$} (N2);
    \draw[->] (I2) -- node[weight, pos=0.28, below right] {$W_2$} (N1);
    \draw[->] (I2) -- node[weight, below] {$W_4$} (N2);

    \draw[->] (N1) -- node[weight, above right] {$W_5$} (N3);
    \draw[->] (N2) -- node[weight, below right] {$W_6$} (N3);

    \draw[->, draw=red!75!black] (N1) to[out=125, in=55, looseness=5.5] node[weight, above=5pt, text=red!75!black] {$W_7$} (N1);

    \draw[->] (N3.east) -- ++(1.1,0) node[right, font=\small] {$y_3$};
    \draw[->, dashed, draw=teal!70!black] (6.4, 1.6) node[above, font=\small, text=teal!70!black] {$y_3^*$} -- (N3.north);

    \draw[->, dotted, thick, draw=teal!70!black] (N3.south) |- (3.2, -2.3) -| node[pos=0.25, below, font=\footnotesize, text=teal!70!black] {Broadcast error $\varepsilon = y_3^* - y_3$ to $\mathcal{C}_{\pi, m}$} (1.6, -1.5);

    
\end{tikzpicture}
\caption{Architecture of the three-neuron benchmark network. Synaptic weights $W_1, \dots, W_6$ and the self-recurrent connection $W_7$ (highlighted in red) are adapted online by decentralized controllers $\mathcal{C}_{\pi, m}$ driven by the broadcast output tracking error $\varepsilon$.}
\label{fig:3neuron_arch}
\end{figure}

Unlike static networks, each synaptic parameter $\theta_m$ exhibits intrinsic continuous-time first-order filtering dynamics discretized at sampling interval $h > 0$:
\begin{equation}
\label{eq:synapse_dyn}
\theta_m(k+1) = (1 + \tau h)\,\theta_m(k) + \gamma h \, u_m(k),
\end{equation}
where $\tau < 0$ is the synaptic decay rate, $\gamma > 0$ is the input gain, and $u_m(k)$ is the control action generated by an autonomous $\mathcal{C}_{\pi, m}$ controller  following~\eqref{eq:control}--\eqref{eq:ts}:
\begin{equation}
\label{eq:synaptic_control}
u_m(k) = \Psi_{k, m} \sum_{j = 1}^k K_{i, m} \left( y_3^*(j) - y_3(j) \right) h,
\end{equation}
with the adaptive modulation sequence $\Psi_{k, m}$ updated via:
\begin{equation}
\Psi_{k, m} = \Psi_{k-1, m} + K_{p, m} \left( K_{\alpha, m} e^{-K_{\beta, m} k} - y_3(k) \right).
\end{equation}

\noindent
In this formulation, no intermediate target values are assigned to the hidden neurons $N_1$ and $N_2$. Instead, the network operates under a \textit{broadcast output-error} paradigm: the reference trajectory $y_3^*(k)$ is provided exclusively at the output neuron $N_3$ ($y_3(k) = z_3(k)$), and the resulting scalar tracking error $\varepsilon(k) = y_3^*(k) - y_3(k)$ is broadcast to all local controllers $\mathcal{C}_{\pi, m}$. Symmetry breaking across the decoupled controllers is achieved through distinct proportional-integral gain scaling ($K_{p,m}, K_{i,m}$) and staggered activation delays. Each controller treats the concurrent updates of all other pathways as unmodeled dynamic disturbances, adjusting its weight online without computing cross-layer Jacobians.
Practically, in a supervised learning context, the input signals $I_1(k)$ and $I_2(k)$ correspond to the input training features, while the reference trajectory $y_3^*(k)$ represents the associated target training outputs.

\subsection{Online Resilience to Abrupt Topological Disruption}

A critical challenge in decentralized systems is maintaining stability and performance under structural degradation, such as, for instance, abrupt connection dropout. To evaluate this resilience, the extended topology incorporating an additional direct synaptic pathway $W_7$ is subjected to a topological perturbation during online tracking.

Specifically, tracking is initialized under the nominal fully connected topology. At a designated iteration step $k_{\text{drop}}$, connection $W_7$ is abruptly severed by forcing: 

\begin{equation}
W_7(k) = 0, \quad \forall k \ge k_{\text{drop}}.
\end{equation}

\noindent
This perturbation induces a sudden loss of transmitted signal flow and shifts the operating point of the output neuron $N_3$. Under the proposed control framework, no offline retraining is invoked. The remaining controllers $\mathcal{C}_{\pi, m}$ perceive the sudden loss of connection $W_7$ purely as an exogenous disturbance step. By virtue of the integral accumulation in~\eqref{eq:synaptic_control} and the adaptive modulation $\Psi_{k,m}$, the remaining controllers automatically recalibrate their synaptic weights online to compensate for the missing signal path, re-establishing asymptotic tracking of $y_3^*(k)$ in real-time.

\subsection{Empirical Results and Performance Evaluation}

To illustrate the empirical efficacy of the proposed decentralized control scheme, numerical simulations were conducted on the benchmark architecture subjected to piecewise-constant shifts in both sensory inputs and target reference levels ($y_3^* \in \{0.55, 0.60\}$). The experiments address two primary questions: (i) Can decoupled synaptic controllers driven solely by a broadcast output error achieve accurate target tracking without global loss gradients? and (ii) Does the closed-loop network possess intrinsic resilience to abrupt topological changes without offline retraining?

\paragraph{Nominal Tracking and Dynamic Convergence}

\begin{figure}[t]
    \centering
    \begin{minipage}[t]{0.48\textwidth}
        \centering
        \vspace{0pt}
        \includegraphics[width=\linewidth]{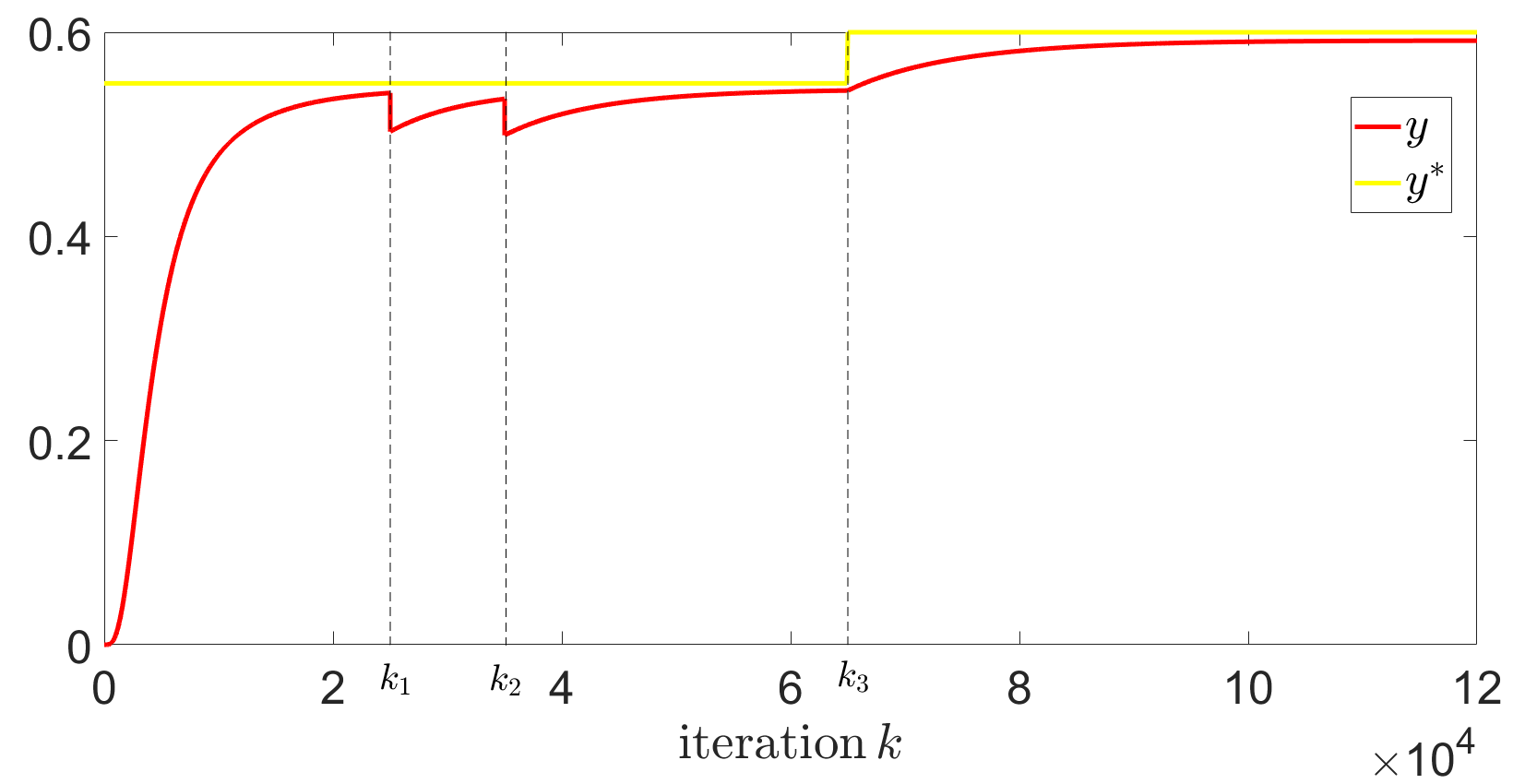}
        \caption*{(a) Nominal output tracking performance of the benchmark architecture: target trajectory $y_3^*(k)$ versus node activation $y_3(k)$.}
        \label{fig:nominal_tracking}
    \end{minipage}
    \hfill
    \begin{minipage}[t]{0.48\textwidth}
        \centering
        \vspace{0pt}
        \includegraphics[width=\linewidth]{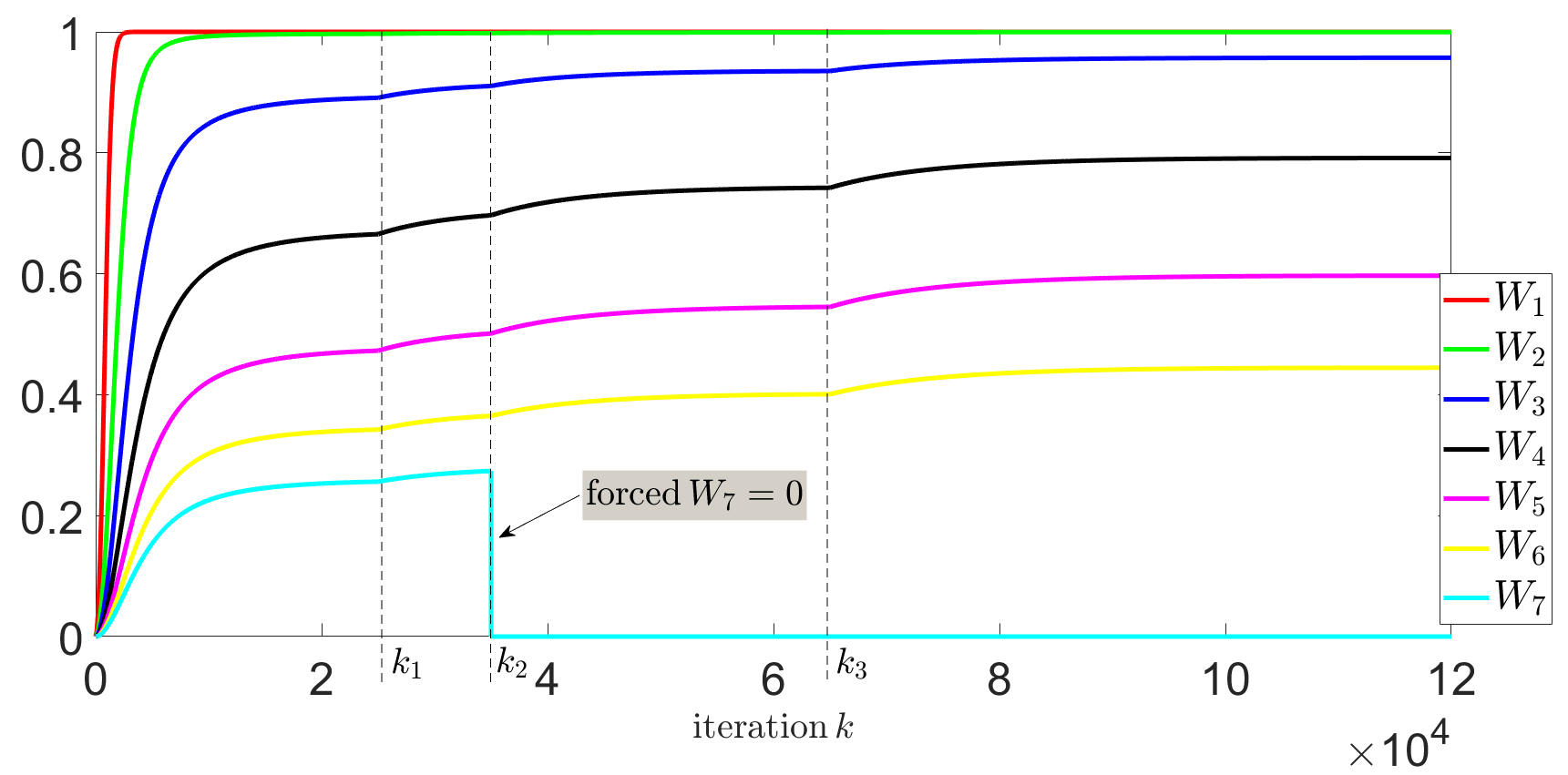}
        \caption*{(b) Dynamic response to connection break at $k = k_{\text{drop}}$: node activation $y_3(k)$ exhibits a transient deflection before regaining convergence onto $y_3^*(k)$.}
        \label{fig:fig_dropout_resilience}
    \end{minipage}
    \caption{Closed-loop tracking performance and topological fault tolerance of the three-neuron benchmark.}
    \label{fig:benchmark_results}
\end{figure}

In the nominal configuration, all synaptic pathways are operational and all parameters $\theta_m(0)$ are initialized at zero without prior calibration. As illustrated in Fig.~\ref{fig:benchmark_results}, the local synaptic weights converge smoothly, steering the output activation $y_3(k)$ rapidly toward the piecewise-constant target trajectory $y_3^*(k)$ across successive input transitions (marked by instants $k_1, k_2, k_3$). Despite the absence of a centralized optimizer or explicit backpropagation paths, the cross-coupling terms between hidden and output units are dynamically absorbed as bounded disturbances by the adaptive integral law~\eqref{eq:synaptic_control}.

\paragraph{Resilience Against Structural Perturbations: Connection Inhibition ($W_7 = 0$)}

To test structural plasticity, the unmodeled topological disturbance is introduced at step $k_2 = k_{\text{drop}}$ by setting $W_7(k) = 0$ for all $k \ge k_{\text{drop}}$.

As shown in Fig.~\ref{fig:benchmark_results}(b), the closed-loop system exhibits a three-phase recovery response:
\begin{enumerate}
    \item \textbf{Transient Deflection:} Immediately following the disconnection of  $W_7$, the output neuron $N_3$ experiences a momentary tracking deviation.
    \item \textbf{Autonomous Dynamic Reallocation:} In response to the resulting broadcast error surge, the decentralized controllers driving the remaining pathways automatically ramp up their control effort, adjusting the remaining synaptic weights $W_m$ ($m \neq 7$) to compensate for the missing signal flux.
    \item \textbf{Asymptotic Recovery:} Without altering controller hyperparameters or invoking offline retraining, the network autonomously restores steady-state tracking of $y_3^*(k)$.
\end{enumerate}

\section{Towards a Dynamic Growing Connectome: The NetPMC Framework}

While the three-node benchmark establishes the local tracking and disturbance rejection properties of the proposed $\mathcal{C}_{\pi}$ controller, scalable neural systems require mechanisms to autonomously reconfigure their wiring. In this section, we outline the generalized architecture of the \texttt{NetPMC} framework, bridging decentralized closed-loop tracking with developmentally growing topological graphs. 
The code is available on \url{https://doi.org/10.5281/zenodo.15072598}.

\subsection{Architectural Formulation and Dual Information Propagation}

\begin{figure}[tb!]
\centering
\includegraphics[width=0.65\linewidth]{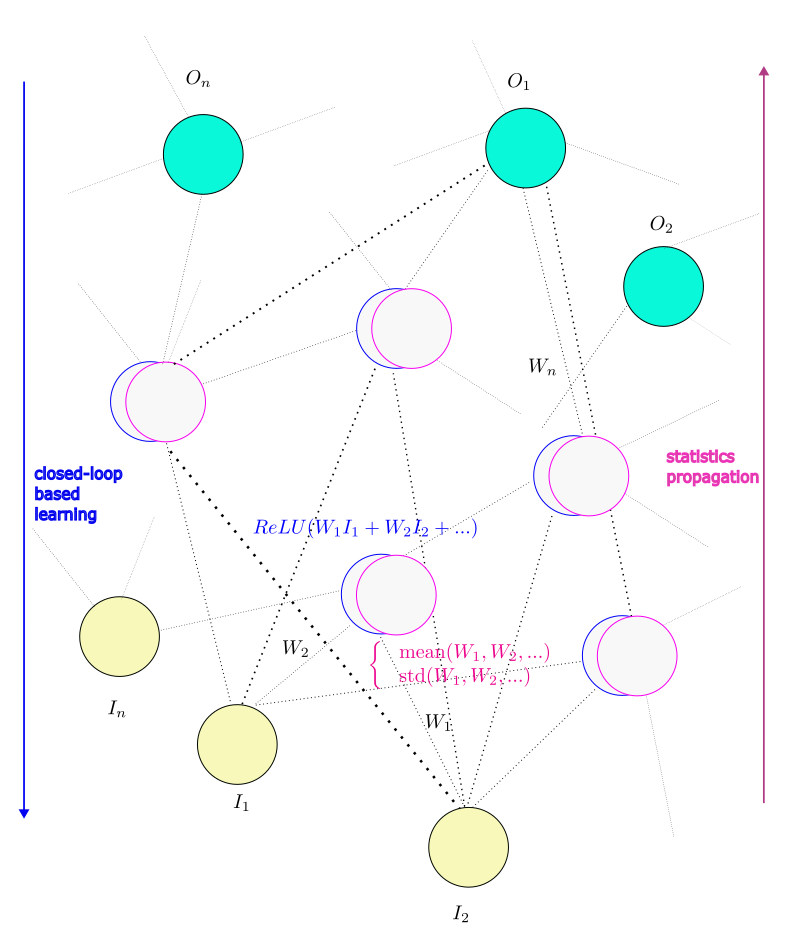}
\caption{Schematic representation of the \texttt{NetPMC} connectome framework. Input units $I_n$ feed an evolving pool of nodes through dynamic synaptic weights $W_{ij}$. Each edge concurrently conveys closed-loop adaptive activation and local statistical descriptors of the wiring, enabling decentralized developmental restructuring.}
\label{fig:connectome_scheme}
\end{figure}

Consider a network comprising $N_{\text{in}}$ inputs, an evolving hidden pool of processing nodes, and $N_{\text{out}}$ output units (see the scheme in Fig.~\ref{fig:connectome_scheme}). The network topology at discrete step $k$ is described by a directed adjacency graph $\mathcal{G}(k) = (\mathcal{V}(k), \mathcal{E}(k))$, where vertices $\mathcal{V}(k)$ denote active nodes and edges $\mathcal{E}(k)$ represent directional synaptic pathways equipped with time-varying transmission weights $W_{ij}(k)$.

A defining feature of the \texttt{NetPMC} framework is the \textit{dual propagation} mechanism operating across each edge $(j, i) \in \mathcal{E}(k)$:
\begin{enumerate}
    \item \textbf{Closed-Loop Dynamic Learning:} Forward neural activations are driven by decentralized controllers $\mathcal{C}_{\pi, ij}$, which adapt the synaptic weights $W_{ij}(k)$ in real-time to minimize output tracking error, as detailed in Section~3.
    \item \textbf{Statistical Wiring Propagation:} Concurrently with forward activation signals, each node aggregates and propagates basic summary statistics (such as the running mean and standard deviation) of its incoming synaptic weights.
\end{enumerate}
By relaying these lightweight local dispersion metrics along processing pathways, downstream nodes gain continuous visibility into upstream wiring activity and convergence stability without requiring a centralized supervisory graph monitor.

\subsection{Continuous Dynamic Learning}

To illustrate the real-time adaptation capability of \texttt{NetPMC} under persistent temporal non-stationarities, we evaluate the network on an online functional tracking task: learning the time-derivative of a continuous oscillatory waveform: $y^*(t) = \frac{d}{dt} \sin(\omega t) = \omega \cos(\omega t)$, where the input nodes receive solely the primal oscillatory signal $x(t) = \sin(\omega t)$. The \texttt{NetPMC} architecture must dynamically discover the phase lead and gain scaling online using only instantaneous output errors.

\begin{figure}[t]
    \centering
    \begin{minipage}[t]{0.48\textwidth}
        \centering
        \vspace{0pt}
        \includegraphics[width=\linewidth]{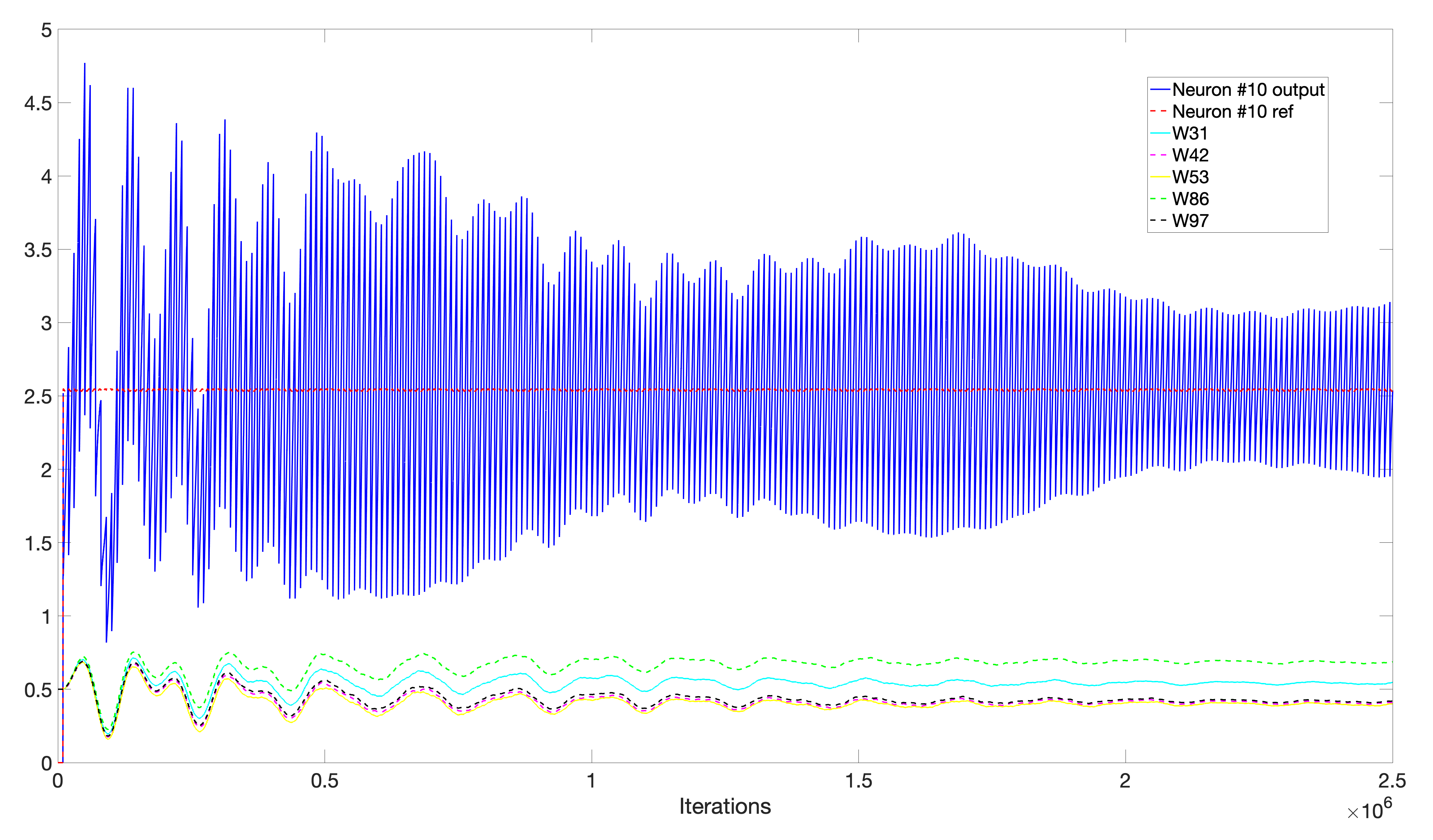}
        \caption*{(a) Top: Amplitude tracking of the derivative reference $y^*(t)$. Bottom: Real-time trajectories of the decentralized synaptic weights $W_{ij}(k)$ adapting without global gradient backpropagation.}
    \end{minipage}
    \hfill
    \begin{minipage}[t]{0.48\textwidth}
        \centering
        \vspace{0pt}
        \includegraphics[width=\linewidth]{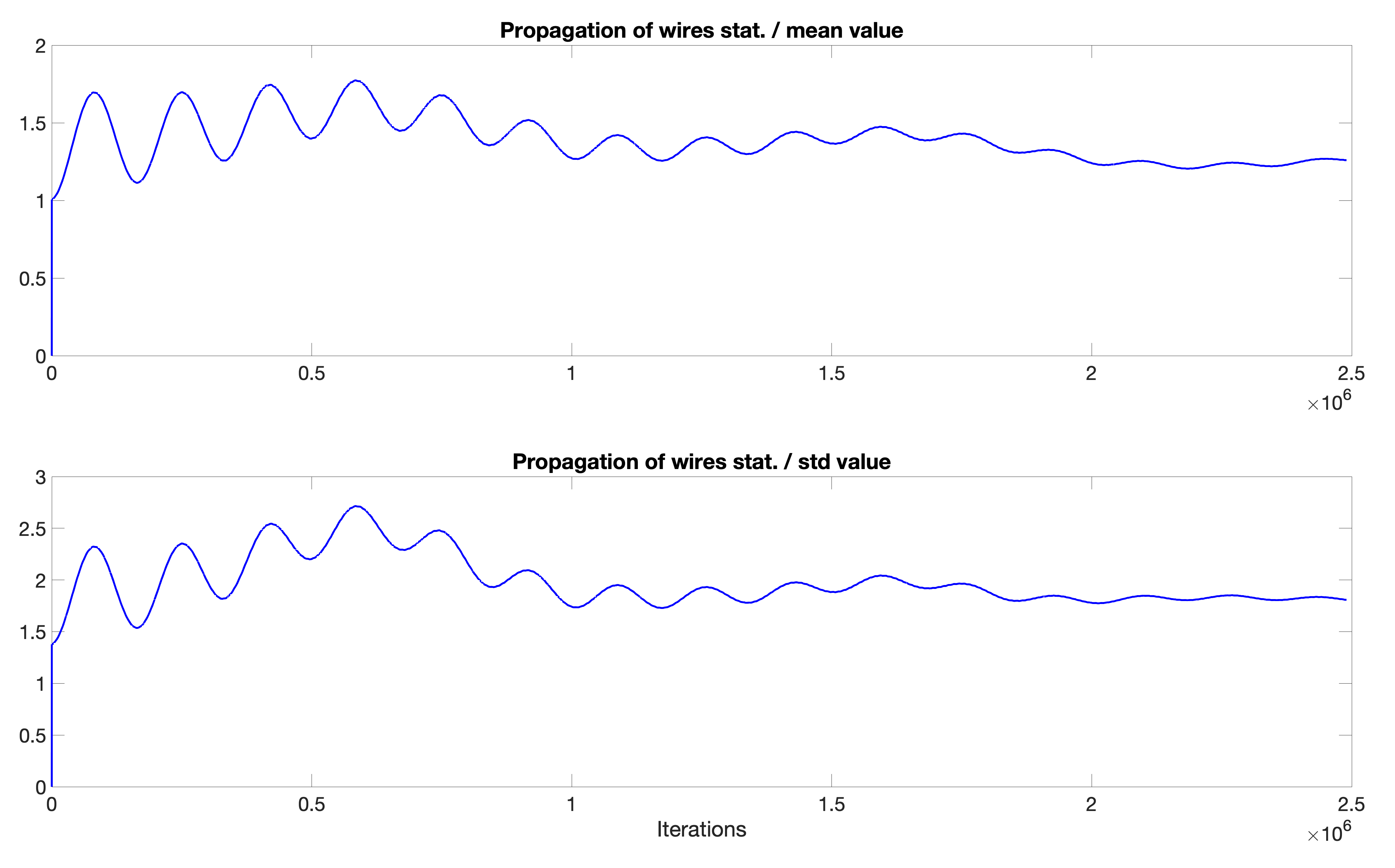}
        \caption*{(b) Evolution of the propagated statistical weight descriptors during the online tracking of the sinusoidal derivative.}
    \end{minipage}
    \caption{Dynamic learning of functional signal variation in \texttt{NetPMC}.}
    \label{fig:Netpmc_results}
\end{figure}

As depicted in Fig.~\ref{fig:Netpmc_results}, the decentralized controllers $\mathcal{C}_\pi$ rapidly tune the synaptic weights $W_{ij}(k)$ across the initial transient cycles. The output node achieves tight trajectory tracking of $y^*(t)$, actively compensating for phase delays inherent to the continuous leak dynamics. Concurrently, the propagated statistical descriptors settle onto bounded steady-state regimes, reflecting the emergence of a coordinated functional pathway within the active connectome.

\subsection{Developmental Plasticity and Architectural Perspectives}

The integration of closed-loop tracking within a dynamically wired graph provides a control-theoretic foundation for developmental machine learning \cite{Kowaliw2014Growing}. Specifically, combining local tracking errors with lightweight wiring statistics opens several direct architectural extensions:
\begin{itemize}
    \item \textbf{Activity-Dependent Synaptogenesis and Pruning:} Synaptic edges exhibiting prolonged near-zero contributions ($W_{ij} \approx 0$) or persistent erratic variance can be dynamically pruned, while new candidate connections are spawned between correlated active clusters.
    \item \textbf{Hierarchical Multi-Task Clustering:} Partitioning large-scale connectomes into functional sub-graphs driven by localized set-points enables modular learning, where distinct node clusters specialize in discrete dynamic sub-tasks while accounting for mutual neuronal interactions.
    \item \textbf{Bidirectional Credit Inference:} Extending the current feedforward control authority to include reciprocal feedback channels will allow local tracking demands at output layers to hierarchically steer upstream topological sprouting.
\end{itemize}

This illustrates how the proposed distributed control formulation imparts intrinsic fault tolerance to the connectome: topological changes are treated as exogenous disturbance steps, rejected online by local feedback mechanisms.

\subsection{Discussion of Observations}

Recent exploratory studies such as \cite{Itoh2024Rethinking} have likewise investigated propagating statistical summaries across neural layers, supporting the relevance of distributed wiring descriptors.

These preliminary numerical outcomes provide an empirical proof-of-concept for gradient-free connectome learning. While standard gradient-based architectures typically require recalculating global Jacobians and executing backward passes upon architectural changes, the proposed closed-loop formulation adapts continuously in real-time. 

Regarding analytical guarantees, recent theoretical breakthroughs \cite{Tamekue1, Tamekue2} have formalized the reachability and small-time boundary control of Hopfield-type RNNs and neural fields using  flow representations, which provide a rigorous mathematical foundation that we envision adapting in future work to formally prove the closed-loop asymptotic convergence of the \texttt{NetPMC} adaptive connectome dynamics.

\section{Conclusion and Perspectives}

This paper presented a preliminary empirical study applying model-free control principles to online neural network training. The numerical simulations illustrate the viability of decentralized trajectory tracking, particularly when the network is subjected to real-time structural perturbations (such as edge removal) and shifting data distributions without requiring explicit gradient backpropagation.   

Building upon this initial study, future work focuses on extending this local control-theoretic paradigm to larger, self-organizing architectures. In particular, this approach forms the foundation of the \texttt{NetPMC} framework (Network Para-Model Control), which generalizes tracking dynamics to evolving connectomes, enabling distributed information processing and dynamic connection pruning.

\subsection*{Generative AI Use}

During the preparation of this work, the authors used ChatGPT (OpenAI) and Gemini (Google) in order to assist with manuscript organization including literature search and English language editing. After using these tools, the authors reviewed and edited the content as needed and take full responsibility for the content of the published article.

\bibliography{l4dc2027-biblio}

\end{document}